\documentclass[aps,twocolumn,nofootinbib,preprintnumbers,groupedaddress]{revtex4-1}
\pdfoutput=1

\usepackage{tabu}
\usepackage[usenames,dvipsnames,table]{xcolor}
\usepackage{graphicx,amsmath,amssymb,amsthm,multirow,array,bm,bbm,esint}
\usepackage[mathscr]{eucal}
\usepackage[bbgreekl]{mathbbol}
\usepackage{epsf,amsfonts}
\usepackage{slashed}
\usepackage{hyperref}
\usepackage{mathdots}

\usepackage{ulem}

\hypersetup{
    pdfstartview={FitH},    % fits the width of the page to the window
    pdftitle={Chaos in AdS2},    % title
    pdfauthor={Felix Haehl and Moshe Rozali},     % author
    colorlinks=true,       % false: boxed links; true: colored links
    linkcolor=blue,          % color of internal links (change box color with linkbordercolor)
    citecolor=red,        % color of links to bibliography
    filecolor=magenta,      % color of file links
    urlcolor=blue           % color of external links
}

\newcommand{\uL}{\hat{u}}

\usepackage{upgreek}
\newcommand{\prop}{\upvarepsilon}
\newcommand{\B[1]}{{\cal B}^{{\scriptscriptstyle(#1)}}}

\usepackage{tikz}

\begin{document}

\title{Fine-Grained Chaos in $AdS_2$ Gravity}

\author{Felix M.\ Haehl}
\author{Moshe Rozali}

\affiliation{Department of Physics and Astronomy, University of British Columbia,\\
6224 Agricultural Road, Vancouver, B.C.\ V6T 1Z1, Canada.}

%%%%%%%%%%%%%%%%%%%%%%%%%%%%%%%%%%%%%%%%%%%
\begin{abstract}
Quantum chaos can be characterized by an exponential growth of the thermal out-of-time-order four-point function up to a scrambling time $\uL_*$. We discuss generalizations of this statement for certain higher-point correlation functions. For concreteness, we study the Schwarzian theory of a one-dimensional time reparametrization mode, which describes $AdS_2$ gravity and the low-energy dynamics of the SYK model. We identify a particular set of $2k$-point functions, characterized as being both  ``maximally braided'' and
``$k$-OTO", which exhibit exponential growth until progressively longer timescales $\uL_*^{(k)} \sim (k-1) \, \uL_*$. We suggest an interpretation as scrambling of increasingly fine-grained measures of quantum information, which correspondingly take progressively longer time to reach their thermal values.
\end{abstract}


 \maketitle

\section{Introduction}

The out-of-time-order (OTO) four-point function $F(\uL) = \langle V(\uL) W(0) V(\uL) W(0) \rangle/(\langle VV \rangle \langle WW \rangle)$ in a thermal state serves as a diagnostic of quantum chaos \cite{Larkin:1969aa,Shenker:2013pqa,Shenker:2013yza,Leichenauer:2014nxa,Maldacena:2015waa,Kitaev:2015aa}. A manifestation of this is the existence of a time regime where the (connected and regularized) part of $F(\uL)$ grows exponentially:\footnote{ Throughout this letter, we we denote Euclidean times as $u$ and real times as $\uL$.} $F(\uL)_{conn.} \sim e^{\lambda_L(\uL-\uL_*)}$. The {\it scrambling time} $\uL_*$ is larger than the typical timescale of thermal dissipation by a factor of the logarithm of the entropy of the system. It has thus been suggested that it quantifies a more fine-grained aspect of thermalization, a process that has been coined {\it scrambling} \cite{Hayden:2007cs,Sekino:2008he,Lashkari:2011yi}.

In this letter we aim to explore generalizations of these statements. We consider higher-point correlation functions in OTO configurations. We will suggest a particular generalization of the four-point chaos correlator, which we call the ``maximally braided'' OTO correlator. As we will see, the maximally braided $2k$-point function is a function of $k$ Lorentzian insertion times and has several interesting features: 
\begin{enumerate}
\item There exist Lorentzian insertion time configurations for which it exhibits exponential growth up until a time $\uL_*^{(k)} \sim (k-1) \uL_*$. These configurations are such that the correlator is {\it maximally OTO}, i.e., they display the highest possible number of switchbacks in real time.
\item The Lyapunov exponent describing the speed of this growth is the same $\lambda_L$ as for the four-point function. The longer time scales are associated with the higher-point correlators being more fine grained quantities, thus can be made progressively smaller initially.
\end{enumerate}
We demonstrate these features in a particular model, which is known to be maximally chaotic (i.e., the Lyapunov exponent is as large as universally allowed in any quantum system, $\lambda_L = \frac{2\pi}{\beta}$ \cite{Maldacena:2015waa,Maldacena:2016hyu,Kitaev:2017awl}): the Schwarzian theory of a single time reparametrization mode, describing the fluctuations of the location of the boundary in $AdS_2$ gravity coupled to scalar matter fields.

\section{OTO Correlation Functions}
\subsection{Backreaction in $AdS_2$}
Our starting point is the calculation of backreaction of matter fields in Euclidean $AdS_2$ space. We follow previous discussions in \cite{Almheiri:2014cka,Maldacena:2016upp,Engelsoy:2016xyb,Jensen:2016pah}, which the reader is invited to consult for further details. The gravitational action reduces to a boundary term, which describes the dynamics of the soft mode $t(u)$:
\begin{equation}
\label{eq:Igrav}
-I_{grav}=\frac{1}{\kappa^2}\int du \, \bigg[-\frac{1}{2}\left(\frac{t''}{t'}\right)^2+\left(\frac{t''}{t'}\right)' \bigg]
\end{equation}
This is the Schwarzian action, which is determined by a pattern of spontaneous and explicit conformal symmetry breaking. The coupling $\kappa$ is our expansion parameter: in gravity it is proportional to $G_N^{1/2}$ (the bulk Newton constant) and it scales as $N^{-1/2}$ in the SYK model.

Note that the SYK model \cite{Kitaev:2015aa,Maldacena:2016hyu} has an additional energy scale $J$, which appears in the gravity calculation as a UV cutoff. The dominance of the soft modes over the massive modes of the SYK model, for certain quantities, stems from those quantities being UV sensitive. We believe this is the case for the special class of correlation functions discussed here, and therefore that the time scales we unravel are also relevant to the SYK model. However, for simplicity we restrict our attention to the purely gravitational calculation, representing the contribution of the soft mode to correlation functions.

We couple the gravity theory to a matter action which represents external massless particles:
\begin{equation}
\label{matter}
-I_{matter}= \frac{1}{2\pi}\int du_1 du_2 \frac{t'(u_1) t'(u_2)}{(t(u_1)-t(u_2))^2}\,j(u_1)j(u_2)
\end{equation}
where  $j$ is a source for the (dimension 1) operator whose correlator we are calculating.

To compute correlators perturbatively in a black hole background, we transform $t(u)=\tan(\frac{\tau(u)}{2})$, corresponding to working with temperature $\beta = 2 \pi$, and expand around the saddle: $\tau(u)=u+\kappa \, \prop(u)$.

To leading order in $\kappa$ the Schwarzian action gives a quadratic term, and hence a propagator for the mode $\prop(u)$. This propagator can be written as
\begin{equation}
\label{prop}
\begin{split}
\langle \prop(u)\prop(0)\rangle&= \frac{1}{2 \pi}\bigg[\frac{
2 \sin \, u - (\pi +u)}{2} \,(\pi+u) \\
&\qquad\qquad\qquad + 2 \pi \Theta(u) (u-\sin \,  u) \bigg]
\end{split}
\end{equation}
where we take the
coefficients $a,b$ appearing in \cite{Maldacena:2016upp} to zero (this corresponds to a gauge choice). Further expansion of the Schwarzian action gives self-interaction  terms for $\prop(u)$, suppressed by factors of $\kappa$. These are required for calculating general correlation functions. We will see that those interactions terms are not needed for our purposes.

Similarly, we can expand the matter action (\ref{matter}). We write the expansion in $\kappa$ as
\begin{equation}
-I_{matter} = \frac{1}{2\pi} \int du_1 du_2 \, \frac{j(u_1)j(u_2)}{4\sin^2 (\frac{u_{12}}{2})} \sum_{p\geq 0} \kappa^p \, {\cal B}^{(p)}(u_1,u_2)
\end{equation}
where $u_{12} \equiv u_1-u_2$.
The leading order contribution comes from  the two-point function in the absence of backreaction. It is the conformal correlator at finite temperature, i.e., $\B[0]= 1$. We will also need the first and second order expansions, corresponding to the way the matter sources the soft mode $\prop(u)$ to orders $\kappa$ and $\kappa^2$ \cite{Sarosi:2017ykf}:
\begin{equation}
\begin{split}
\B[1](u_1,u_2)&= \prop'(u_1)+\prop'(u_2)-\frac{\prop(u_1)-\prop(u_2)}
{\tan(\frac{u_{12}}{2})} \\
\B[2](u_1,u_2)&= \frac{1}{4 \sin^2(\frac{u_{12}}{2})} \Big[ (2+\cos u_{12})\, (\prop(u_1)-\prop(u_2))^2 \\
& + 4 \sin^2 \big(\frac{u_{12}}{2}\big)\,\prop'(u_1)\prop'(u_2) \nonumber\\
& - 2  \,\sin  u_{12}  \,\left( \prop(u_1) - \prop(u_2) \right) \left( \prop'(u_1)+\prop'(u_2)\right) \Big]
\end{split}
\label{Bfactor}
\end{equation}
In order to compute a Euclidean $2k$-point function up to ${\cal O}(\kappa^n)$, one has to sum the relevant diagrams arising from this expansion: first, one writes all possible products of $k$ instances of $\B[p_i](u_{2i-1},u_{2i})$, which are relevant at $n$-th order in perturbation theory (i.e., $\sum_i p_i \leq n$). In this product, one then contracts $\prop$'s either with propagators \eqref{prop}, or with higher-point vertices arising from expanding the action \eqref{eq:Igrav} to higher orders in $\kappa$. This quickly gets complicated (see appendix \ref{app:sixpt} for examples). We will now present a particularly interesting class of observables for which this task simplifies considerably.

\subsection{Systematics of the Calculation}
\label{sec:conventions}

Consider coupling the Schwarzian theory, describing gravity in $AdS_2$ space, to $k$ distinguishable matter fields representing the coupling to external operators $V_i$ with $i=1,...,k$. Our aim is to calculate $2k$-point correlation functions involving the operators $V_1(u_1), V_1(u_2), \ldots , V_k(u_{2k -1}) ,V_k(u_{2k})$. We proceed as follows: $(1)$ We calculate the Euclidean correlators. Without loss of generality, for each pair of insertions of the same operator, say $V_i(u_{2i-1})$ and $V_i(u_{2i})$, we order the Euclidean times as $u_{2i-1}>u_{2i}$. The remaining relations between Euclidean insertion times determine the order in which the operators occur in the correlation function. $(2)$ Then, to discuss Lorentzian times we analytically continue by setting $u_r \rightarrow \delta_r + i  \uL_r$ for all $r=1,\ldots,2k$. We then analyze the late time dependence on Lorentzian times $\uL_r$. $(3)$ Ultimately we are interested in putting equivalent operators at coincident Lorentzian times, $\uL_{2i-1} = \uL_{2i}$. The short time regulators $\delta_r$ (which are ordered in the same way as the original Euclidean times) serve to regulate the divergence in this limit. We write below terms at leading order in $\delta_{ij} \equiv \delta_i - \delta_j$, which are universal in the sense that they contain the exponential behavior we are interested in (see the discussion in \cite{Roberts:2014ifa}).

We start by discussing the computation of Euclidean correlators.
The Euclidean time ordering determines the ordering of operators in the correlator. We are interested in a specific set of orderings, which we will call {\it maximally braided} correlators, for which the calculation becomes particularly simple. To describe those correlators we need to introduce some conventions.

The backreaction calculation involves in intermediate steps Heaviside $\Theta$-functions, resulting from the propagator of the soft mode \eqref{prop}. Organizing these will be crucial. We choose to write all step functions canonically as $\Theta(u_i-u_j)$ with $i>j$, using 
$\Theta(x)=1-\Theta(-x)$. We then use the configuration of these step functions to uniquely characterize the different possible operator orderings of the correlation function. For example, the time ordered correlator $\langle V_1(u_1) V_1(u_2) \cdots V_k(u_{2 k -1}) V_k(u_{2k}) \rangle$, with the canonical ordering $u_1 > u_2> \ldots > u_{2k}$, is characterized as being the term in the generic Euclidean $2k$-point function with no step functions.

Since we are interested in the exponential growth in the chaos regime, we will only keep terms that are dominant at late times. Longest living modes can be characterized as a coefficient in the generic Euclidean correlator with the maximum number of step functions. It is simpler to evaluate, and subtracting off all other time orderings does not influence the information we are interested in.

\subsection{Maximally Braided Correlator} 

\begin{figure}
\includegraphics[width=.28\textwidth]{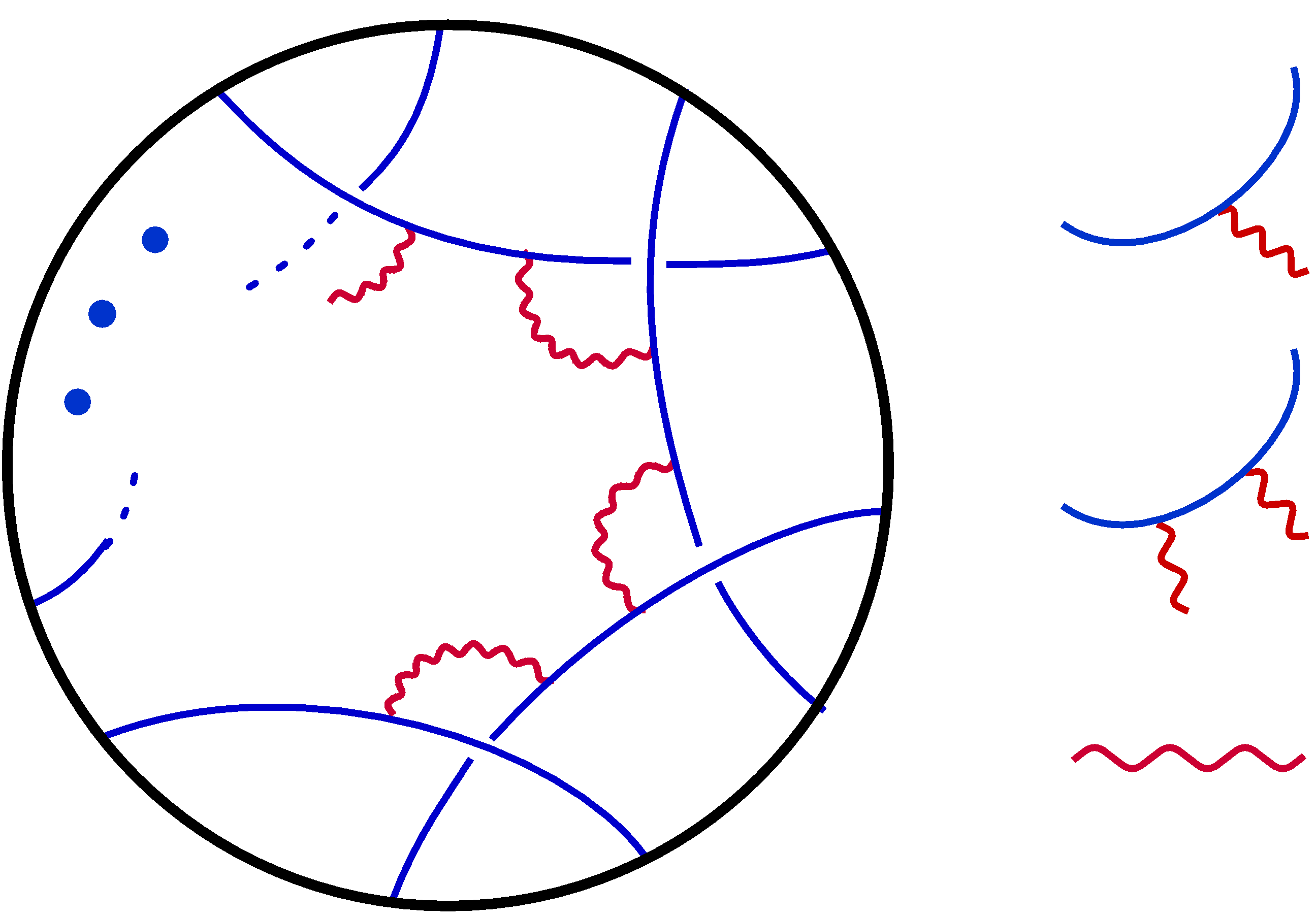}
\put(-139,13){$1$}
\put(-70,0){$2$}
\put(-102,-7){$3$}
\put(-50,18){$5$}
\put(-42,40){$4$}
\put(-50,71){$7$}
\put(-68,93){$6$}
\put(-97,100){$9$}
\put(-123,93){$8$}
\put(-153,30){$2k$}
\put(-150,40){$\vdots$}
\put(-137,83){$\iddots$}
\put(0,80){$\;= \; \B[1](u_{2i-1},u_{2i})$}
\put(0,46){$\;= \; \B[2](u_{2i-1},u_{2i})$}
\put(0,15){$\;= \; \langle \prop \prop \rangle$}
$\qquad\qquad\qquad$
\caption{General maximally braided $2k$-point correlator (first term obtained by expanding out commutators in $F_{2k}$): only diagrams of the type shown contribute to the connected correlator $F_{2k}$ at leading order in $\kappa$. The arrangement of insertions along the circle indicates the ordering in Euclidean time.}
\label{general}
\end{figure}
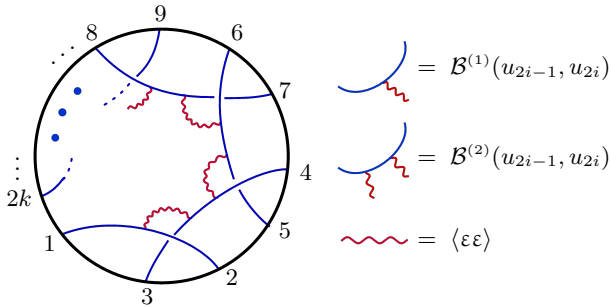

 Our subtracted maximally braided correlator can be characterized by the appearance of precisely $k-1$ step functions, ``braiding" every pair of operators with the consecutive pair. Elementary combinatorics shows that this characterization is equivalent to computing a product of commutators (see appendix \ref{app:simplifications} for details). We thus define our basic observable of interest as
\begin{widetext}
\begin{equation}
\label{eq:F2kdef}
  F_{2k}(u_1,\ldots,u_{2k}) = \frac{ \big{\langle} V_1(u_1) [V_2(u_3), V_1(u_2)]\, [V_3(u_5), V_2(u_4)]\, [V_4(u_7), V_3(u_6)]  \cdots [V_{k}(u_{2k-1}), V_{k-1}(u_{2k-2})] V_{k}(u_{2k}) \big{\rangle} }{ \langle V_1(u_1) V_1(u_2) \rangle \cdots \langle V_k(u_{2k-1}) V_k(u_{2k}) \rangle} 
\end{equation}
\end{widetext}
The maximally braided configuration is obtained by dropping all commutator brackets (see Fig.\ \ref{general}). The commutators in $F_{2k}$ serve to subtract subleading pieces from the maximally braided configuration. $F_{2k}$ is then just the coefficient of a term in the generic Euclidean correlator with $k-1$ step functions.
 We argue below that to leading order in perturbation theory, $F_{2k}$ can be computed using only the Feynman diagrams of the type illustrated in Fig.\ \ref{general}.

Note that thus far we are discussing the Euclidean time ordering, or equivalently the operator ordering in the correlator. This determines the combinatorics of the calculation, and is the source of the simplification we exploit. We discuss the independent issue of the Lorentzian time ordering, which is crucial to the understanding of the different time scales involved in the correlator we compute, further below.

\subsection{Example: OTO Four-Point Function}

Consider the correlator $\langle V_1(u_1)[V_2(u_3),V_1(u_2)]V_2(u_4) \rangle$. We demonstrate here the simplified calculation that picks out this particular combination (which describes precisely the dominant term in the chaos regime), without the need to calculate the full Euclidean or Lorentzian 4-point function. We then generalize that process for higher-point functions.

Using the simplifications described in appendix \ref{app:simplifications}, we compute $F_4$ as the coefficient of $\Theta(u_{32})$ in the exchange of a soft mode between two bilinears:
\begin{equation}
\label{2OTO}
\begin{split}
F_4 &=\kappa^2 \, \langle \B[1](u_1,u_2) \B[1](u_3,u_4)\rangle\big{|}_{\Theta(u_{32})}  + {\cal O}(\kappa^3)  \\
& =   \left\{ \frac{4\,\kappa^2}{\delta_{12}\delta_{34}}\, [ (u_{23}-\sin u_{23}) ] + {\cal O}(\delta_{ij}^{-1}) \right\} + {\cal O}(\kappa^3)
\end{split}
\end{equation}
Introducing $\delta_{ij}\equiv \delta_i - \delta_j$, we have already used the benefit of hindsight and extracted the leading divergence as $\delta_{ij}\rightarrow 0$ for the analytic continuation $u_r \rightarrow \delta_r + i \uL_r$ with $\uL_1=\uL_2$, $\uL_3 = \uL_4$. We can thus complete the analytic continuation to the OTO chaos region by simply setting $u_{23} \rightarrow  i \uL_{23}$.\footnote{Note that $\Theta(u)= \frac{1}{2\pi i} \int d\omega \, \frac{e^{i \omega u}}{w- i \epsilon^+}$ depends only on the real part of $u$. In our context this means the step functions are sensitive to the operator ordering, but not to the Lorentzian time ordering.} The term $\sin u_{23}$ in (\ref{2OTO}) then gives an exponentially growing term $e^{\lambda_L |\uL_2-\uL_3|}$, 
with $\lambda_L =1 =\frac{2 \pi}{\beta= 2 \pi}$, as expected. The time scale associated to this exponential growth, where the correlator becomes of order one, is the {\it scrambling time} $\hat{u}_* \sim \log(\kappa^{-2}) \sim \log(G_N^{-1}) \sim \log(N)$, or with units: $\uL_* \sim \frac{\beta}{2\pi} \, \log(\frac{2\pi}{\beta \kappa^2}) $. 

Indeed, \eqref{2OTO} is the result obtained by evaluating the full 4-point function, specializing to the operator ordering $\langle V_1(u_1)V_2(u_3)V_1(u_2)V_2(u_4) \rangle$, subtracting off the time-ordered part, and expanding in small $\delta_{ij}$ (c.f.\ \cite{Maldacena:2016upp}).
 
Note that the exponentially growing factor is associated with one soft mode propagator, relating the even and odd parts of two matter perturbations. We see below that such pattern persists for higher-point correlators, where one such exponential factor is associated with any exchange of operators relative to the canonical ordering. Any such exchange is reflected by the presence of a (canonically ordered) step function we use to organize the calculation. Each such step function is accompanied by a similar propagator factor and hence by an exponentially growing mode. This is the basic structure of the results derived below.

 \section{Higher-Point Correlators}
 
 Consider the six point function $F_6$ as defined in \eqref{eq:F2kdef}, following the process outlined and demonstrated in the previous section. The combination $\langle V_1(u_1) [V_2(u_3), V_1(u_2)] [V_3(u_5), V_2(u_4)] V_3(u_6)\rangle $ is obtained from the generic Euclidean six-point function by isolating the terms involving the product of step functions $\Theta(u_{32}) \Theta(u_{54})$. We claim that the necessary presence of this product of step functions specifies a unique diagram that can contribute to the (connected and subtracted) maximally braided correlator, to leading order in $\kappa$.

Indeed the diagram depicted in Fig.\ \ref{general} (for $k=3$) contains the minimal ingredients necessary to produce the two step functions defining the maximally braided ordering we are interested in. Such diagram is of order $\kappa^4$. Other diagrams of the same order, for example disconnected ones or those involving a 3-point self-interaction of the soft mode, will have fewer step functions. They contribute only to other correlators, where the braiding is less than maximal, or get subtracted off in the combination $F_6$. Similarly, diagrams involving more than two $\prop$-propagators contribute to $F_6$ but at higher orders in $\kappa$.

We are therefore faced with the relatively easy calculation of the following contribution to Fig.\ \ref{general}:
\begin{equation}
\label{eq:F6}
F_6
= \kappa^4 \, \langle \B[1](u_1,u_2) \, \B[2](u_3,u_4)\,\B[1](u_5,u_6)\rangle \big{|}_{\Theta(u_{32})\Theta(u_{54})}
\end{equation}
up to corrections of ${\cal O}(\kappa^5)$.
Since we will eventually set $u_r = \delta_r + i \uL_r$, we can further use the simplifications of appendix \ref{app:simplifications}. The result to leading order in $\kappa$ and to leading order in the regulators $\delta_{ij}$ is
\begin{equation}
\label{eq:F6res}
\begin{split}
F_6&\sim\frac{24\,\kappa^4}{\delta_{12}\delta_{34}^2\delta_{56}}\, (u_{23}-\sin u_{23})(u_{45}-\sin u_{45}) 
\end{split}
\end{equation}
In appendix \ref{app:sixpt} we illustrate how to calculate the full six-point function and reproduce this simple result for the maximally braided subtracted correlator.

The calculation of the eight-point function is similar. To leading order in $\kappa$ and $\delta_{ij}$ we find:
\begin{equation}
\label{eq:F8res}
\begin{split}
F_8%& = \kappa^6 \, \langle \B[1](u_1,u_2) \, \B[2](u_3,u_4)\, \B[2](u_5,u_6)\,\B[1](u_7,u_8)\rangle \\
    &\sim \frac{144\,\kappa^6}{\delta_{12}\delta_{34}^2\delta_{56}^2\delta_{78}}\, \prod_{i=1}^3\; (u_{2i,2i+1} - \sin u_{2i,2i+1})
\end{split}
\end{equation}

Similar results are obtained for higher order maximally braided correlators $F_{2k}$. Those continue to obey the pattern evident from extrapolating \eqref{eq:F6res} and \eqref{eq:F8res}.

\section{Lorentzian Times}
\label{sec:lorentzian}

We now turn to the analytic continuation $u_r \rightarrow \delta_r + i \uL_r$ in more detail. Our assumptions so far concerned Euclidean time ordering and the first term in $F_{2k}$ (dropping all commutators) corresponds to the choice $\delta_1 > \delta_3 > \delta_2 > \delta_5 > \ldots$. The late time growth indicating quantum chaos is, however, sensitive to the ordering of real Lorentzian times $\uL_r$. As we will now see, there is an independent way to characterize the real time ordering of the correlator. The {\it proper-OTO number} of $F_{2k}$ is determined by the real time ordering and it affects both the associated Lyapunov exponents and the associated scrambling time scales. We will see that the correlator we discuss involves the time scale $\hat{u}_*$, but also longer time scales, depending on  the proper-OTO number.

\subsection{Types of OTO Correlators}

Our maximally braided correlators involve $k$ swaps of neighbouring operators as compared to the canonical (time ordered) configuration. It also has the distinguishing feature that it can  be  {\it maximally OTO}: its analytic continuation allows for configurations that are as much out of time order as any $2k$-point function can be.

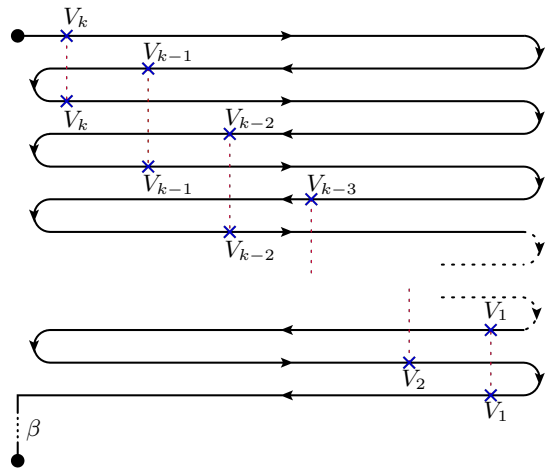
\begin{figure}
\includegraphics[width=.4\textwidth]{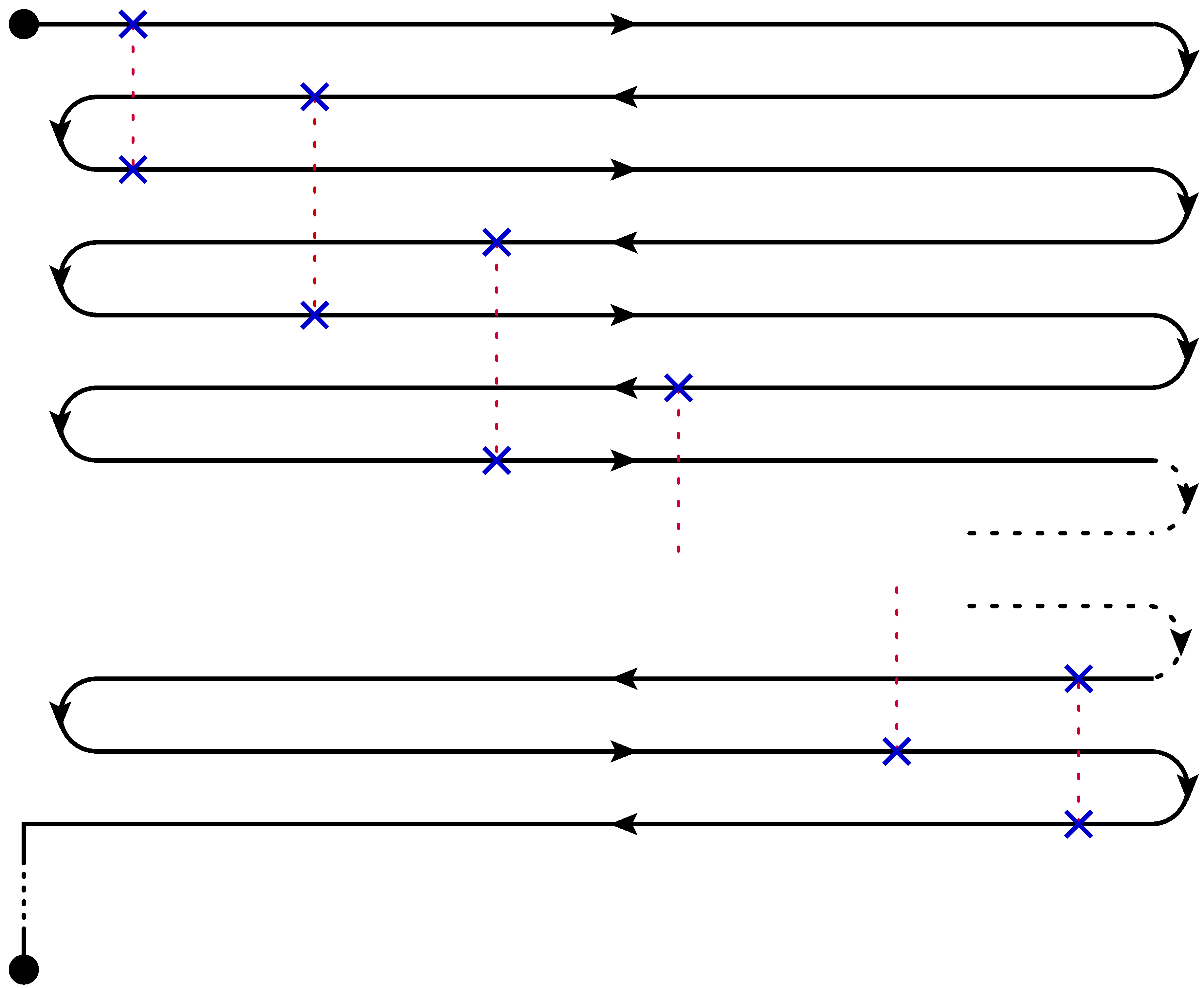}
\put(-183,168){$V_k$}
\put(-183,129){$V_k$}
\put(-153,155){$V_{k-1}$}
\put(-153,105){$V_{k-1}$}
\put(-122,130){$V_{k-2}$}
\put(-122,80){$V_{k-2}$}
\put(-91,105){$V_{k-3}$}
\put(-55,31){$V_{2}$}
\put(-24,19){$V_{1}$}
\put(-24,57){$V_{1}$}
\put(-197,12){$\beta$}
\caption{Complex time contour representation of the maximally braided correlator. We show the first (and dominant) term in the expansion of commutators in $F_{2k}$. Lorentzian time runs horizontally. We depict the Lorentzian time configuration which is maximally OTO. Operators are separated by small imaginary times, which enforces the operator ordering along the contour.}
\label{contour}
\end{figure}

The {\it proper-OTO number} indicates the minimal number of switchbacks in the complex time contour that is required to represent a correlator \cite{Haehl:2017qfl}. The proper-OTO number of a $2k$-point function is at most $k$. In the case of $F_{2k}$, maximal OTO number is achieved by the real time ordering $\uL_1=\uL_2 > \uL_3 = \uL_4 > \ldots > \uL_{k-1} = \uL_k$, which we focus on. The associated contour is shown in Fig.\ \ref{contour}. Most other configurations of real times lead to a smaller proper-OTO number (i.e., the correlator can be represented on a contour with fewer switchbacks).  We now show the significance of this characterization of the possible Lorentzian time orderings of our correlators.

\subsection{Time Scales}

Let us now discuss the physical significance of the proper-OTO number. Using the result from the previous section, we have the following behavior for real time separations $|\uL_{2i} - \uL_{2i+1}| \gg  1 =\frac{\beta}{2\pi}$:\footnote{ This assumption about separation of time scales is simply to extract a clear exponential time dependence from the trigonometric factors in $F_{2k}$. Dropping it would allow transient effects where the time dependence interpolates between different regimes. Note that this assumption is mild: the time scale $\frac{\beta}{2\pi}$ is sometimes referred to as the {\it dissipation time} over which, e.g., two-point functions decay. It is parametrically smaller than the scrambling time scales we are interested in.}

\begin{equation}\label{eq:F2exp}
\begin{split}
  F_{2k} &\sim {\cal N}\, \frac{\exp \left( \sum_{i=1}^{k-1}|\uL_{2i} - \uL_{2i+1}|-(k-1)\, \uL_* \right)}{\delta_{12}\delta_{34}^2 \cdots \delta_{2k-3,2k-2}^2\delta_{2k-1,2k}} 
\end{split}
\end{equation}
with scrambling time $\uL_* \sim \log (\kappa^{-2})$, associated with the growth of the 4-point function. 
The normalization ${\cal N}$ is ${\cal O}(1)$ and has an alternating sign depending on the sign of $\uL_{2i}-\uL_{2i+1}$. 
Note
the appearance of the term $(k-1)\,
\uL_{*}$ in the exponent, reflecting the fact that the connected $2k$-point functions are proportional to $\kappa^{2 (k -1)}$.

Depending on the real time ordering, the connected $2k$-point function $F_{2k}$ exhibits different growth patterns  as function of different time separations. We focus on the proper $k$-OTO configurations: these are maximally OTO, i.e $\uL_{2i} > \uL_{2i-1}$ for all $i$. The time differences in the exponent in \eqref{eq:F2exp} are then all positive and cancel telescopically (recalling that we set $\uL_{2i} = \uL_{2i-1}$ for all $i$), yielding $F_{2k} \sim e^{\uL_1 - \uL_{2k-1} - (k-1) \uL_*}$.

Thus the correlator in this case is a function of a single time separation $\uL_{1,2k-1}$, corresponding to a measurement which is only sensitive to the total duration of the experiment. Despite being scrambled in different ``channels", the chaotic growth of $F_{2k}$ does not saturate after the scrambling time $\uL_*$ and continues unabated until $\uL_{1,2k-1}$ reaches the {\it $k$-scrambling time} 
\begin{equation}
\uL_*^{(k)} \sim (k-1) \uL_* \,.
\end{equation}
The Lyapunov exponent for this growth is still $\lambda_L^{(k)} = 1 = \frac{2\pi}{\beta}$, but the longer time scale is associated with our chosen correlators being sensitive to more fine grained quantum chaos: they start off smaller and continue to grow for a longer time.

Let us now discuss briefly configurations with less than maximal OTO-number. For example, proper $(k-1)$-OTO configurations are obtained by swapping the order of a single pair of real times, say $\uL_1$ and $\uL_3$,  giving a correlator which can be represented on a contour with only $k-1$ switchbacks.  The exponents in \eqref{eq:F2exp} do not quite add up anymore, and we get $F_{2k} \sim e^{2\uL_3 - \uL_1 - \uL_{2k-1} - (k-1) \uL_*}$. There is now a two-dimensional space of time dependence on both $\uL_{31}$ and $\uL_{3,2k-1}$. If, e.g., $1\ll \uL_{31}\ll \uL_*$, then after a total duration of the experiment $\uL_\text{tot} = \uL_{3,2k-1} \sim (k-2) \uL_*$ the observable $F_{2k}$ already reaches size of ${\cal O}(1)$.  %In appendix \ref{ReducedOTO}, we show that in a particular limit the $(k-1)$-OTO correlator can be reduced to a $2(k-1)$-point correlator, which exhibits exponential growth until time  $(k-2) \uL_*$. 
Working recursively, we see that less than maximal OTO configurations can exhibit intermediate time scales and transient behavior. It would be interesting to explore this in more detail.

\section{Discussion}

We have argued that there exists new physically interesting data in higher-point out-of-time-order (OTO) correlation functions. These are qualitatively similar to the OTO four-point function used to diagnose quantum chaos. However, the observables $F_{2k}$ we constructed in \eqref{eq:F2kdef} display an exponential growth for a longer time $\uL_*^{(k)} \sim (k-1) \, \uL_*$. That is, there exists a hierarchy of timescales associated with scrambling, probed by increasingly fine grained (OTO) observables. 

This is reminiscent of similar hierarchies encountered in the context of unitary $k$-design in quantum circuit complexity \cite{Roberts:2016hpo,Cotler:2017jue}. It would be interesting to explore this connection. Similarly, it would be an intriguing task to explore the experimental relevance, or the precise operational meaning of the hierarchy of $k$-scrambling times (for instance, along the lines of \cite{Halpern:2016zcm,Halpern:2017abm}). An interpretation in terms of echo experiments,  or more theoretically as quantifying operator growth  by the size of repeated commutators, seem possible.

Several other questions immediately spring to mind: It would be interesting to repeat the calculation in the Lorentzian setting, as a variant fo the standard shock wave calculation \cite{Shenker:2013pqa,Shenker:2013yza,Stanford:2014jda} (one would have to interpret the maximal braiding in that context). Similarly, one would like to make precise the connection to the formalism of \cite{Mertens:2017mtv}. Extensions to higher dimensions (e.g. \cite{Gu:2016oyy}) and exploration of butterfly velocities would be interesting, for example in the context of 2-dimensional CFTs at large central charge \cite{Roberts:2014ifa}. It is also interesting to explore whether those $k$-OTO  correlators obey some bounds along the lines of \cite{Maldacena:2015waa} (see also \cite{Tsuji:2017fxs}).

Finally, we hope to explore other types of $2k$-point OTO correlators, such as the (suitably regularized) ``tremolo'' correlator $\langle (W(t) V(0))^k \rangle$. This might shed light on the physical significance of abstract arguments about the structure of OTO correlators \cite{Haehl:2017qfl,Haehl:2017eob}.

%%%%%%%%%%%%%%%%%%%%%%%%%%%%%%%%%%%%%%%%%%%%%%%%%%%%%
\begin{acknowledgments}
%%%%%%%%%%%%%%%%%%%%%%%%%%%%%%%%%%%%%%%%%%%%%%%%%%%%
We thank Ahmed Almheiri, Pawel Caputa, Nicole Yunger Halpern, Kristan Jensen, Rob Myers, Dan Roberts, Brian Swingle and Beni Yoshida for helpful discussions. FH is grateful for hospitality by University of California, Santa Barbara, where part of this work was done. FH is supported through a fellowship by the Simons Collaboration `It from Qubit'. MR is supported by a Discovery grant from NSERC.
\end{acknowledgments}

\begin{widetext}

\appendix

\section{Technical simplifications}
\label{app:simplifications}

We collect here some simplifications that make the evaluation of $F_{2k}$ more efficient in practice. 
The generic Euclidean $2k$-point function is invariant under permutations of the time arguments, but most terms in it multiply  some step functions. Any particular choice of Euclidean time ordering singles out some of these terms. Using the conventions of section \ref{sec:conventions}, the fully time ordered correlator has no  step functions. Any step function signals the exchange of two insertion times with respect to the above defined canonical ordering.
A correlator with a single step function would be one with a single pair of neighbouring operators exchanged relative to the time ordered one, e.g., $\langle V_1(u_1) V_2(u_3) V_1(u_2) V_2(u_4) \cdots \rangle$. 
Based on general arguments, the time ordered correlators reach their thermal value much faster than the scrambling times we are interested in. Similarly, any term without a sufficient number of step functions has time ordered pieces in it, which factorize off and decay in the chaos regime. As an example, observe that $\langle V_1V_2V_1V_2V_3V_3 \rangle \sim \langle V_1V_2V_1V_2\rangle \langle V_3V_3 \rangle$ for relative time differences $\uL_{ij} \gg 1$. We are thus interested in the term with the maximal number of step functions. 

The observable $F_{2k}$ is constructed precisely such that one starts with the maximally braided configuration $\langle V_1(u_1) V_2(u_3) V_1(u_2) V_3(u_5) V_2(u_4) V_4(u_7) V_3(u_6)  \cdots V_{k}(u_{2k-1}) V_{k-1}(u_{2k-2}) V_{k}(u_{2k}) \rangle$
(illustrated in Figs.\ \ref{general} and \ref{contour}) and then subtracts off all the pieces which contribute to it, but only involve a lower number of step functions. This ensures that we compute the leading term in the chaos regime, but nothing else. We extract the part of the generic Euclidean $2k$-point correlator with precisely $k-1$ step functions of the form $\Theta(u_{32})\Theta(u_{54}) \cdots \Theta(u_{2k-1,2k-2})$, which is nothing but $F_{2k}$. 

For instance, the computation of $F_6$ at ${\cal O}(\kappa^4)$ can be illustrated as follows:
\begin{equation}
\begin{split}
  F_6 \big{|}_{{\cal O}(\kappa^4)} &= \frac{1}{ \langle V_1V_1 \rangle \langle V_2 V_2 \rangle \langle V_3 V_3 \rangle} \big( \langle V_1 V_2 V_1 V_3 V_2 V_3 \rangle - \langle V_1 V_2 V_1 V_2 V_3 V_3 \rangle - \langle V_1 V_1 V_2 V_3 V_2 V_3 \rangle + \langle V_1V_1V_2V_2V_3V_3 \rangle \big) \big{|}_{{\cal O}(\kappa^4)}\\
  &= \begin{gathered}\includegraphics[width=.58\textwidth]{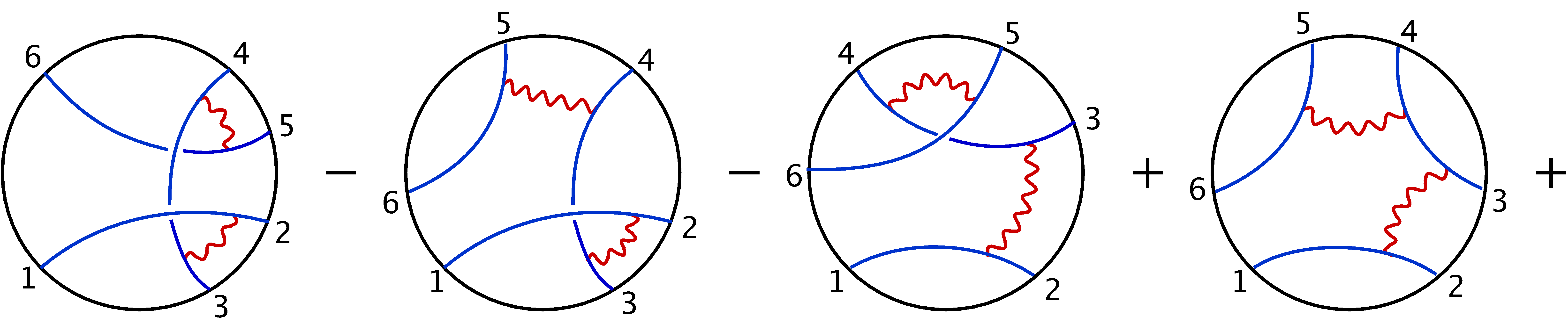}\end{gathered}
   \; \text{other } \prop\text{-contractions}
 \end{split}
 \label{eq:F6diagrams}
\end{equation}
Of interest to us are coefficients in the full (unordered) Euclidean six-point function of $\Theta(u_{32})\Theta(u_{54})$, $\Theta(u_{32})$, $\Theta(u_{54})$, and $1$ (no step function). All four contribute to the first diagram; the second and fourth contribute to the second diagram; the third and fourth contribute to the third diagram; and only the term with no step function contributes to the last diagram. In total, only the coefficient of $\Theta(u_{32})\Theta(u_{54})$ contributes to the signed sum \eqref{eq:F6diagrams}. This coefficient contains the maximum number of growing modes in the chaos regime. This is summarized in \eqref{eq:F6}.

Some more simplifications are useful for efficiently computing $F_{2k}$ for large values of $k$. In computing this term, we can anticipate that eventually $\tan(\frac{u_{ij}}{2})\approx \frac{\delta_{ij}}{2}$, where $\delta_{ij} \equiv \delta_i - \delta_j$. Within $F_{2k}$ we can then write $\B[1](u_{2i-1},u_{2i}) \rightarrow \B[1]_{odd}(u_{2i-1})+\B[1]_{even}(u_{2i})$, with
\begin{equation}\label{eq:simpl1}
\begin{split}
\B[1]_{odd}(u_{2i-1})&= \prop'(u_{2i-1}) - \frac{2\prop(u_{2i-1})}{\delta_{2i-1,2i}} \,,\qquad
\B[1]_{even}(u_{2i})= \prop'(u_{2i}) + \frac{2\prop(u_{2i})}{\delta_{2i-1,2i}} 
\end{split}
\end{equation}
etc.. This source for the soft mode separates into two terms, each depending on either $u_{2i-1}$ or $u_{2i}$ but not both.
The second order bilinear $\B[2](u_{2i-1},u_{2i})$ simplifies in a similar way: in that case we only need to keep terms involving both $\prop(u_{2i-1})$ and $\prop(u_{2i})$. Any term involving the square of only one of them would not be able to produce the consecutive step functions $\Theta(u_{2i-1,2i-2})$ and $\Theta(u_{2i+1,2i})$.

Finally, we wish to extract the term involving $k-1$ factors $\Theta(u_{2i+1,2i})$ for $i=1,\ldots,k-1$. This can arise in our perturbation theory only if subsequent bilinears $\B[p_i](u_{2i-1},u_{2i})$ are connected by propagators as in Fig.\ \ref{general}. We only need to retain the part of the soft mode propagator (\ref{prop})  that contains the corresponding step function. We can therefore define a truncated propagator
\begin{equation}\label{eq:simpl3}
\langle \prop(u)\prop(0)\rangle_\text{trunc.} =\Theta(u) \, (u-\sin \,  u) 
\end{equation}
which will be sufficient for computing $F_{2k}$ at leading order in $\kappa$.

\section{The Full Six-Point Function}
\label{app:sixpt}

In the main text, we computed maximally braided $2k$-point functions using an approximation scheme appropriate for extracting the late time growth characteristic of quantum chaos. As a consistency check on our approximations, in this appendix we elaborate on the exact Euclidean six-point function $G_E^{(6)}$ to see how it truncates to \eqref{eq:F6res}.

The leading order diagram is ${\cal O}(\kappa^0)$ and can be represented as
\begin{equation}
  G_E^{(6)} \big{|}_{{\cal O}(\kappa^0)} =\;  \begin{gathered}\includegraphics[width=.095\textwidth]{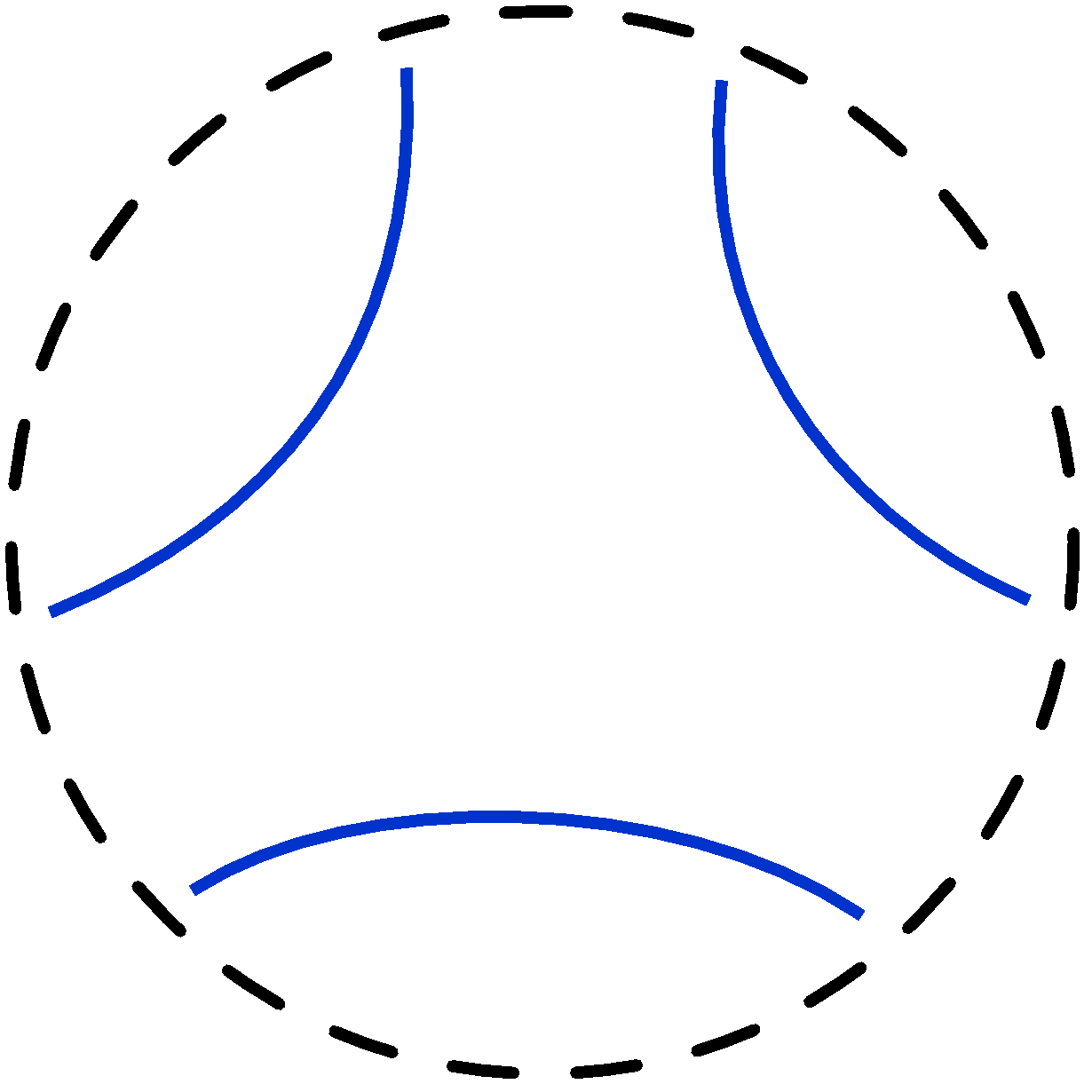}\end{gathered}
\end{equation} 
The dashed external circle indicates that the Euclidean time ordering has not been fixed and one should sum over permutations of the external insertion points (corresponding to ``braiding'' the blue lines). The above contribution is, of course, the completely disconnected product of three two-point functions. 

The next order in perturbation theory involves one $\langle \prop\prop\rangle$ propagator:
\begin{equation}
  G_E^{(6)} \big{|}_{{\cal O}(\kappa^2)} =\;  \begin{gathered}\includegraphics[width=.27\textwidth]{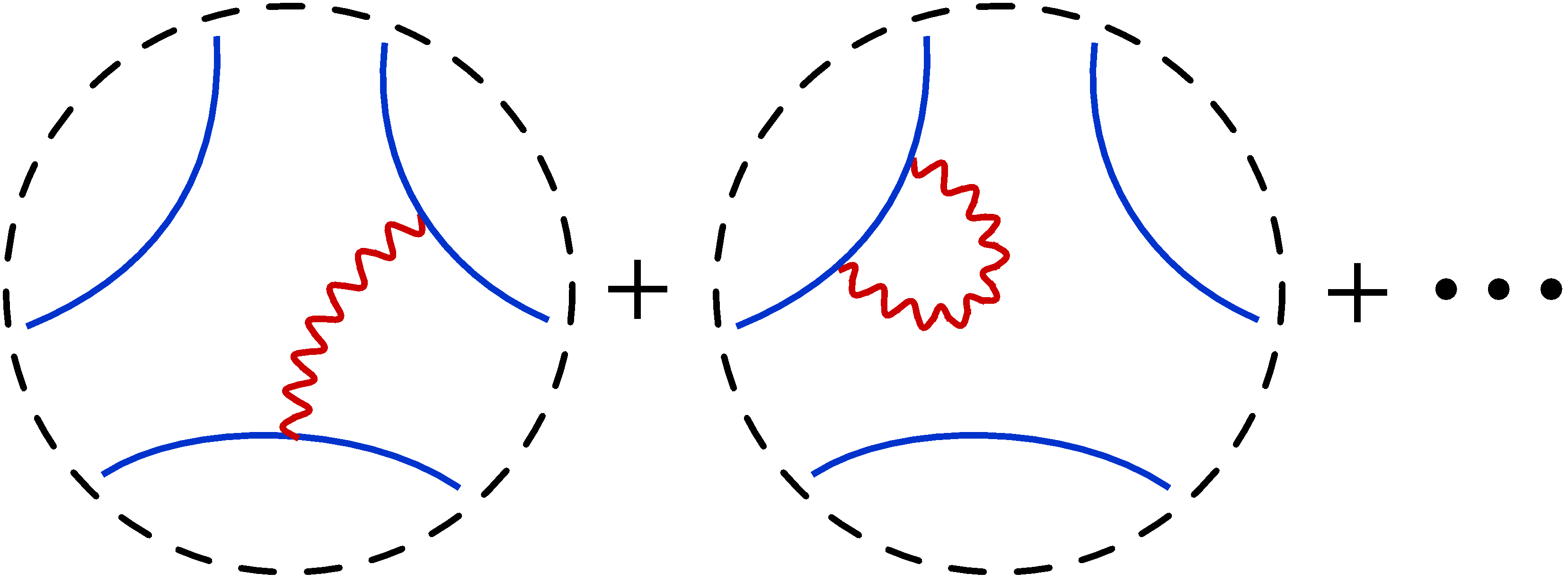}\end{gathered}
\end{equation} 
These contributions factorize in a product of a two-point and a four-point function. 

The contribution of interest to us is the ${\cal O}(\kappa^4)$ part since this involves connected pieces for the first time (which are the ones measuring the $k$-scrambling time scales):
\begin{equation}
  G_E^{(6)} \big{|}_{{\cal O}(\kappa^4)} =\;  \begin{gathered}\includegraphics[width=.7\textwidth]{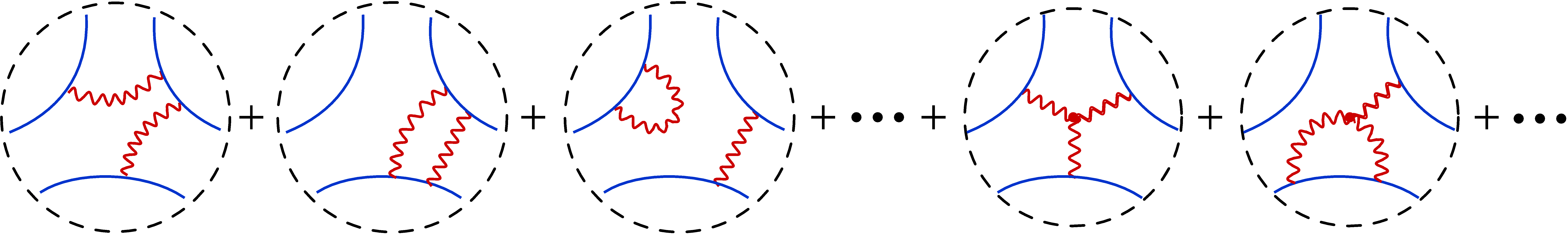}\end{gathered}
\end{equation} 
where we are omitting a few more diagrams which are similar. In the main text, we computed only the first type of diagram since this is the only one contributing to the (subtracted) maximally braided six-point function $F_6$. The remaining diagrams are of the same order in perturbation theory, but do not contain enough step functions to contribute to $F_6$. If we were interested in the generic (unordered) Euclidean correlator, we would have to consider the full set of diagrams. For any specific ordering, one can identify a subset of diagrams that contributes. 

We can indeed compute the diagrams shown above for arbitrary Euclidean time ordering.\footnote{ We have also confirmed this result using the computational method developed in \cite{Almheiri:2014cka}. That calculation is, of course, isomorphic, but in practice somewhat different to implement.} The result is complicated and unilluminating. But from the above diagrams one can readily see which diagrams contribute given a particular combination of step functions.

Higher orders in $\kappa$ are not our concern here, but some such loop calculations for two- and four-point functions have been done in \cite{Maldacena:2016upp}.

%%%%%%%%%%%%%%%%%%%%%%%%%%%%%%%%%%%%%%%%%%%%%%%

\end{widetext}
 \bibliographystyle{apsrev}
 \bibliography{OTOreferences}

\end{document}